\documentclass[12pt,preprint]{aastex}
\usepackage{graphicx, amsmath, amssymb}
\usepackage[dvips]{color}

\newcommand{\pow}[2]{\ensuremath{\mbox{#1}^{\rm #2}}}
\newcommand{\subs}[2]{\ensuremath{\mbox{#1}_{\rm #2}}}
\newcommand{\del}[1]{\ensuremath{\Delta #1}}
\newcommand{\Blos}{\ensuremath{B_{\rm los}}}
\newcommand{\water}{H$_2$O}
\newcommand{\methanol}{CH$_3$OH}
\newcommand{\kmS}{km \pow{s}{-1}}
\newcommand{\cm}[1]{\pow{cm}{#1}}    
\newcommand{\mG}{\pow{mG}{-1}}

\newcommand{\vLSR}{\ensuremath{v_\text{LSR}}}
\newcommand{\src}{M8E}

\newcommand{\Msun}{\subs{M}{\odot}}

\shortauthors{Sarma \& Momjian}
\shorttitle{EVLA Zeeman Effect Detection in the 36 GHz CH$_3$OH maser}

\begin{document}

\title{Detection of the Zeeman Effect in the \\ 
	36 GHz Class I CH$_3$OH maser line with the EVLA}

\author{A.~P.\ Sarma\altaffilmark{1}, E.\ Momjian\altaffilmark{2}}

\altaffiltext{1}{Physics Department, DePaul University, 
2219 N. Kenmore Ave., Byrne Hall 211, 
Chicago IL 60614; asarma@depaul.edu}

\altaffiltext{2}{National Radio Astronomy Observatory, Socorro NM 87801}

\begin{abstract}
We report the first detection of the Zeeman effect in the 36 GHz Class I \methanol\ maser line. The observations were carried out with 13 antennas of the EVLA toward the high mass star forming region M8E. Based on our adopted Zeeman splitting factor of $z = 1.7$ Hz \mG, we detect a line of sight magnetic field of $-31.3 \pm 3.5$ mG and $20.2 \pm 3.5$ mG to the northwest and southeast of the maser line peak respectively. This change in sign over a 1300 AU size scale may indicate that the masers are tracing two regions with different fields, or that the same field curves across the regions where the masers are being excited. The detected fields are not significantly different from the magnetic fields detected in the 6.7 GHz Class II \methanol\ maser line, indicating that \methanol\ masers may trace the large scale magnetic field, or that the magnetic field remains unchanged during the early evolution of star forming regions. Given what is known about the densities at which 36 GHz \methanol\ masers are excited, we find that the magnetic field is dynamically significant in the star forming region.
\end{abstract}


\section{INTRODUCTION}
\label{sINTRO}

Magnetic fields likely play an important role in the star formation process (e.g., \citealt{tt2008}, and references therein). The nature of this role is not yet clear, however, primarily due to the scarcity of observational data on magnetic fields in star forming regions. The Zeeman effect remains the most direct method for measuring magnetic field 
strengths (e.g., \citealt{rmc99}). Observations of the Zeeman effect in H I and 
OH thermal lines have revealed the strength of the magnetic field in the lower 
density envelopes of molecular clouds (e.g., \citealt{bt01}). 
Observations of the Zeeman effect in \water\ masers, on the other hand, offer a window into magnetic fields in the highest density regions ($n \sim 10^9~\cm{-3}$) of star forming regions (\citealt{sarma2008}; \citealt{vdv06}). 

Interstellar masers, being compact and intense, are extremely effective probes of young star forming regions. Recently, \citet{wv2008} carried out the first systematic study of the Zeeman effect in the 6.7 GHz methanol maser line. Such (6.7 GHz) masers are examples of Class II methanol masers, which are known to be probes of the early phases of high mass star forming regions. Yet another class of masers, namely Class I methanol masers, may probe even earlier phases of star forming regions (\citealt{pp2008}). This appears to be the case for M8E, which is a high mass star forming region located at $l = 6\arcdeg.05$, $b = -1\arcdeg.45$, at a distance of 1.5 kpc (\citealt{scf1984}; \citealt{gg76}). M8E is comprised of a bright infrared source (M8E-IR), and a radio \ion{H}{2}\ region about 8$\arcsec$ to its northwest (\citealt{scf1984}; \citealt{dw2009}). Low spatial resolution CO observations by \citet{wright77} showed that M8E is embedded in a dense molecular core. At higher resolution, \citet{mhs92} mapped a bipolar outflow in the $^{12}$CO emission line (\vLSR = 11 \kmS), with blue and red wings spread over 50 \kmS. Based on their VLTI observations, \citet{linz2009} concluded that M8E-IR contains a central protostar of mass 10-15 \Msun.

In this paper, we report the \textsl{first} detection of the Zeeman effect in the 36 GHz Class I \methanol\ maser line. The observations were carried out with 13 antennas of the Expanded Very Large Array (EVLA) toward the high mass star-forming region M8E, and represent an early success of the EVLA. In \S\ \ref{sODR}, we present details of the observations and reduction of the data. The analysis involved in extracting magnetic field information from the Zeeman effect is given in \S\ \ref{sANAL}. The results are presented and discussed in \S\ \ref{sR}.   

\section{OBSERVATIONS \& DATA REDUCTION}
\label{sODR}

Observations of the $4_{-1}-3_0$ methanol maser emission line at 36 GHz were 
carried out using the Expanded Very Large Array (EVLA)
of the NRAO\footnote{The National Radio Astronomy 
Observatory (NRAO) is a facility of the National Science Foundation 
operated under cooperative agreement by Associated Universities, Inc.}
in two 2 hr sessions on 2009 July 9 and 25. Thirteen EVLA antennas equipped 
with the 27$-$40 GHz (Ka-band) receivers were used in these observations. To 
avoid the aliasing that affects the lower 0.5 MHz of the bandwidth for EVLA 
data correlated with the VLA correlator, the spectral line was centered in 
the second half of the 1.56 MHz wide band. The source 3C286 (J1331$+$3030) 
was used to set the absolute flux density scale, while the compact source 
J1733$-$1304 was used as an amplitude calibrator.

The editing, calibration, Fourier transformation, deconvolution, and 
processing of the data were carried out using the Astronomical Image Processing 
System (AIPS) of the NRAO. After applying the amplitude gain corrections of 
J1733$-$1304 on the target source \src, the spectral channel with the strongest 
maser emission signal was split, then self-calibrated in both phase and amplitude 
in a succession of iterative cycles. The final phase and amplitude solutions were 
then applied on the full spectral-line $uv$ data set, and Stokes $I$ and $V$ image 
cubes were made. In the channel with the maser line peak, we recovered about 
70\% of the flux density reported in the single dish observations of \citet{pp2008}.
Further processing of the data, including magnetic field estimates, 
was done using the MIRIAD software package.

\section{ANALYSIS}
\label{sANAL}

For cases in which the Zeeman splitting \subs{\del{\nu}}{z}\
is much less than the line width \del{\nu}, the magnetic field
can be obtained from the Stokes $V$ spectrum, which exhibits a
{\em scaled derivative} of the Stokes $I$ spectrum (\citealt{hgmz93}). 
Here, consistent with AIPS conventions,
$I$ = (RCP$+$LCP)/2, and $V$ = (RCP$-$LCP)/2; RCP is right- and 
LCP is left-circular polarization incident on the antennas, where
RCP has the standard radio definition of clockwise rotation of the 
electric vector when viewed along the direction of wave propagation. 
Since the observed $V$ spectrum may also contain a scaled replica 
of the $I$ spectrum itself, the Zeeman effect can be measured by 
fitting the Stokes $V$ spectra in the least-squares sense to the  equation   
\begin{equation}
\mbox{V}\ = \mbox{aI}\ + \frac{\mbox{b}}{2}\ 
\frac{\mbox{dI}}{\mbox{d}\nu}
\label{eVLCO}
\end{equation}
(\citealt{th82}; \citealt{skzl90}). The fit parameter $a$ is usually the result of small calibration errors in RCP versus LCP, and is expected to be small. In these observations, $a$ was of the order of $10^{-4}$ or less. While eq.~(\ref{eVLCO})\ is strictly true only for thermal lines, numerical solutions of the equations of radiative transfer (e.g., \citealt{nw92}) have shown that it gives reasonable values for the magnetic fields in masers also. In eq.~(\ref{eVLCO}), the fit parameter $b = zB$\,cos\,$\theta$, where $z$ is the Zeeman splitting factor (Hz \mG), $B$ is the magnetic field, and $\theta$ is the angle of the magnetic field to the line of sight (\citealt{ctg93}). The value of the Zeeman splitting factor $z$ for \methanol\ masers is very small, because \methanol\ is a non-paramagnetic molecule. Following the treatment of \citet{wv2008} for the Zeeman splitting of 6.7 GHz methanol masers, we derive the Zeeman splitting factor using the Land$\acute{e}$ $g$-factor based on laboratory measurements of 25 GHz methanol masers (\citealt{jen51}), and find $z$ = 1.7 Hz \mG. While this will likely introduce a systematic bias into the magnetic field results, it remains the best possible estimate pending laboratory measurements of the 36 GHz maser line.

\section{RESULTS \& DISCUSSION}
\label{sR}

\begin{figure}
\centering
\plottwo{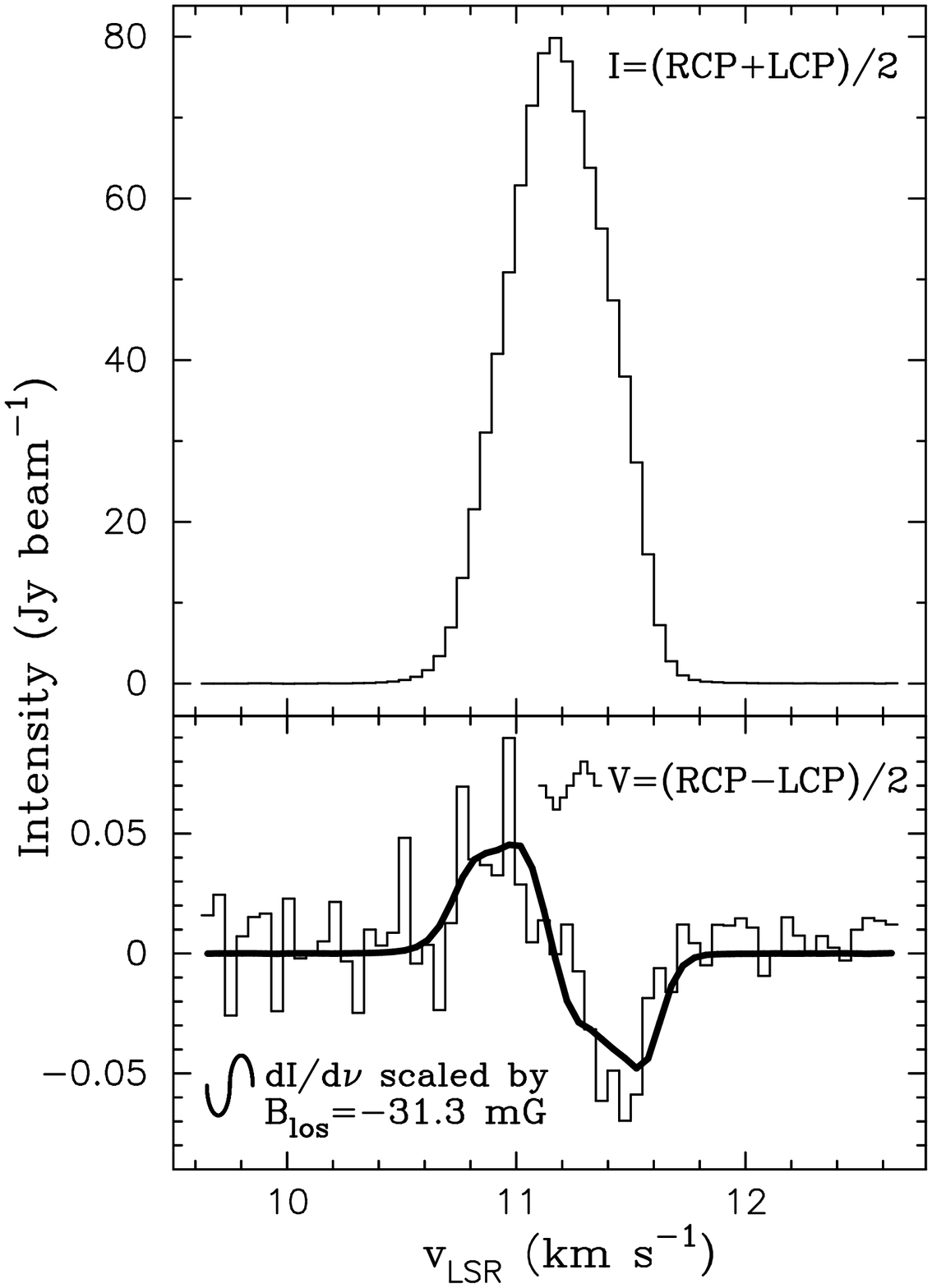}{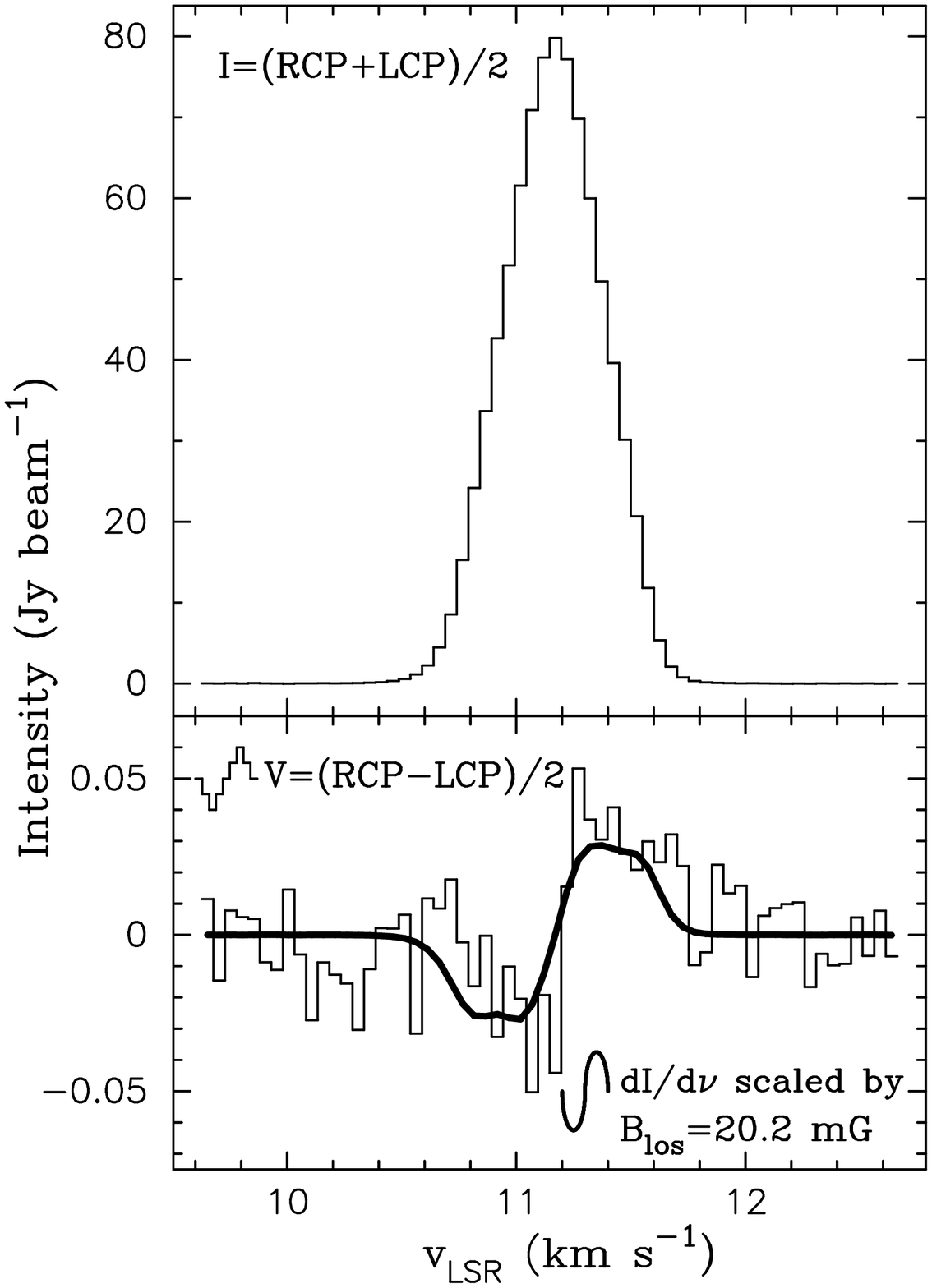}
\caption{Stokes $I$ ({\em top$-$histogram}) and $V$ ({\em bottom$-$histogram})
profiles of the maser toward the (a) northwest and (b) southeast of the maser line peak.
The curve superposed on $V$ in each of the lower frames is the derivative of $I$ scaled 
by a value of \Blos\ = $-31.3 \pm 3.5$ mG in (a), and \Blos\ = $20.2 \pm 3.5$ mG in (b). \label{fIVD}}
\end{figure}

In order to demonstrate our detection of the Zeeman effect in the 36 GHz Class I \methanol\ 
maser line toward M8E, we display the Stokes $I$ and $V$ profiles toward two positions in this source in Fig.~\ref{fIVD}; the two positions are to the northwest and southeast of the maser line peak. As described in \S\ \ref{sANAL}, we determined magnetic fields by fitting the Stokes $V$ spectra in the least-squares sense using equation (\ref{eVLCO}). The values of the fit parameter $b$ (see eq.~\ref{eVLCO}) obtained from this fit are $b = -53.2 \pm 6.0$ Hz for the northwest position, and $b = +34.4 \pm 5.9$ Hz for the southeast position. Using the Zeeman splitting factor $z = 1.7$ Hz \mG\ discussed in \S\ \ref{sANAL}\ above, this gives \Blos\ = $-31.3 \pm 3.5$ mG for the northwest position, and $20.2 \pm 3.5$ mG for the southeast position. By convention, a negative value for \Blos\ indicates a field pointing toward the observer. 

The line-of-sight magnetic field has opposite signs at the two positions for which Stokes $I$ and $V$ profiles are shown in Fig.~\ref{fIVD}, with values of $-$31.3 mG and 20.2 mG, respectively. The observed change in the sign of \Blos\ at these two positions, together with a slight asymmetry in the maser line profiles at each position, indicates that we are observing at least two masers that are very close in position and velocity. The masers are marginally resolved in our C-configuration observations, otherwise the opposite magnetic fields would sum to zero. EVLA B- or A-configuration observations will be necessary to fully resolve the maser components. Such observations will not be possible until late 2010 or early 2011. The observed change in sign of \Blos\ occurs over a size scale of 0$\arcsec$.9, equal to 1300 AU (assuming the distance to M8E is 1.5 kpc). This may mean that the clumps where the 36 GHz maser is being excited come from two different regions where the field is truly different. Alternatively, it may mean that the field lines curve across the region in which the masers are being excited, so that the line-of-sight field traced by one maser is pointed toward us, whereas that traced by the other maser is pointed away from us.

Since this is the first detection of the Zeeman effect in the 36 GHz \methanol\ maser line, we will address the issue of beam squint, a topic which seems to arise every time a Zeeman detection is mentioned. Beam squint arises because the RCP and LCP beams have slightly different pointing centers. Therefore, if a velocity gradient is present across an extended source, beam squint can give rise to a false Zeeman pattern. Since masers are compact sources, beam squint should not be an issue here. Moreover, since the signs of the detected field are reversed at the two positions, it would be difficult to manufacture a velocity gradient that would reverse the signs of the fields as observed.

Our detected \Blos\ values in the 36 GHz Class I \methanol\ maser line toward M8E are similar to the values obtained in the 6.7 GHz Class II \methanol\ maser line observed by \citet{wv2008} toward other sources. \citet{wv2008} detected significant magnetic fields in 6.7 GHz methanol masers with the 100 m Effelsberg telescope in 17 sources, with an average value of 23 mG. Since Class I and Class II \methanol\ masers trace different spatial regions, this may indicate that the magnetic field is the same over the extent of these two regions, giving rise to the possibility that \methanol\ masers trace the large-scale magnetic field in the star forming region. Alternatively, if Class I masers occur in the very early stages of star formation (before the formation of an ultracompact \ion{H}{2}\ region), and Class II masers occur later on, the similarity in \Blos\ for these two classes, if true, may indicate that the magnetic field strength remains the same during the early stages of the star formation process. Certainly, more observations are necessary, especially in regions where both Class I and Class II masers are known to occur, in order to address the possibilities noted above.

Finally, we use our observed value for the magnetic field to compare the magnetic and dynamical energies in these masing regions. The magnetic energy density is given by $3 \Blos^2/8\pi$. Here, we have used the relation $B^2 = 3 \Blos^2$ that was obtained by \citet{rmc99} on statistical grounds. For \Blos\ $\approx$ 25 mG from these observations, the magnetic energy density will be $8 \times 10^{-5}$ ergs \cm{-3}. The kinetic energy density (thermal and turbulent) is given by $(3/2)mn\sigma^2$, where
$\sigma = \Delta v/(8 \mathrm{ln} 2)^{1/2}$ is the velocity dispersion, the mass $m = 2.8 m_p$ (assuming 10\% He; $m_p$ is the proton mass). Since masers occur only in special directions along which they have developed the required velocity coherence, the velocity dispersions in the masing region may be greater than that traced by masers. If we use $\Delta v = 5$ \kmS\ --- a reasonable value (see, e.g., \citealt{cfm99}), we get a kinetic energy density equal to $3 \times 10^{-8}$ ergs \cm{-3}\ for $n = 10^5$ \cm{-3}, the density quoted in \citet{pp2008} at which maser action in the 36 GHz line is maximized. Moreover, suppose we find fields of similar strength in the 44 GHz \methanol\ maser line in a future observation, 44 GHz being the other prominent Class I \methanol\ maser line. Then, considering the density $n = 10^6$ \cm{-3}\ at which maser action in the 44 GHz line is maximized (\citealt{pp2008}), we will get a kinetic energy density equal to $3 \times 10^{-7}$ ergs \cm{-3}. Either way, the magnetic energy is dominant by at least two orders of magnitude, indicating that the magnetic fields must be dynamically significant in these regions.

\section{CONCLUSIONS}
\label{sCONC}
We have detected for the first time the Zeeman effect in the 36 GHz Class I \methanol\ maser line toward the high mass star forming region M8E. The detected values for the line-of-sight magnetic field are $-31.3 \pm 3.5$ mG and $20.2 \pm 3.5$ mG. There may be systematic bias in these values due to the assumed Zeeman splitting factor of $z = 1.7$ Hz \mG, which is based on laboratory experiments on the 25 GHz \methanol\ maser line. Our detected values are similar to the magnetic fields detected in 6.7 GHz Class II \methanol\ masers by \citet{wv2008}. This suggests that \methanol\ masers may trace the large-scale magnetic field; alternatively, the magnetic field may remain the same during the early evolution of the star forming region between the time when Class I and Class II masers are excited. Our detected values for \Blos\ imply that the magnetic field is dynamically significant to these regions. 

Our detection gives rise to a host of questions that need further observational and theoretical work. We are planning future observations of the Zeeman effect in more 36 GHz masers and at higher resolution. We are also planning for observations of 44 GHz masers with the aim of detecting the Zeeman effect. As mentioned above, the 44 GHz line constitutes the other prominent Class I \methanol\ maser line. We are also planning to carry out high angular resolution observations of the Zeeman effect in regions containing both Class I and Class II masers, in order to compare the strength of magnetic fields between the regions traced by these classes. From theorists, we seek detailed models for Class I masers in order to understand their properties better, especially the range of densities at which they are excited. Also, there is a pressing need for laboratory studies of the Zeeman splitting coefficient for \methanol\ masers.  The resolution of some or all of these issues will open up yet another important window into the early stages of high mass star forming regions.

\acknowledgments
This work has been partially supported by a Cottrell College Science Award
(CCSA) from Research Corporation to A.P.S.\ We thank Tom Troland for some
very productive discussions, and an anonymous referee for helpful suggestions. 
We would also like to acknowledge Joseph Booker, a DePaul undergraduate 
student, whose Matlab fitting programs (developed as 
part of a summer project) proved to be extremely useful in cross-checking 
our results. We have used extensively the NASA Astrophysics Data 
System (ADS) astronomy abstract service, and the astro-ph web server. 

\clearpage



\begin{deluxetable}{lccccccrrrrrr}
\tablenum{1}
\tablewidth{0pt}
\tablecaption{Parameters for EVLA Observations \protect\label{tOP}}   
\tablehead{
\colhead{Parameter}  &  
\colhead{Value}}
\startdata
Observation Dates & 2009 July 9 \& 25 \\
Configuration &   C \\
R.A.~of field center (J2000) & $18^\text{h} 04^\text{m} 53^\text{s}.3$ \\
Decl.~of field center (J2000) & $-$24$\arcdeg$26$\arcmin$42$\arcsec$.0  \\
Total Bandwidth & 1.56 MHz \\
No. of channels & 256 \\
Channel Spacing  & 0.051  \kmS \\
Time on Source  & 2 hr \\
Rest Frequency & 36.16929 GHz \\
Velocity at band center\tablenotemark{a} & 13.7 \kmS \\
Target source velocity & 11.2 \kmS \\
Hanning Smoothing &  No \\
FWHM of synthesized beam & $1.76\arcsec \times 0.58\arcsec$ \\
& P.A. $= -7.80\arcdeg$ \\
Line rms noise\tablenotemark{b} &   18 mJy beam$^{-1}$
\enddata
\tablenotetext{a}{The line was centered in the second half of the 1.56 MHz band in
order to avoid aliasing (see \S\ \ref{sODR}).}
\tablenotetext{b}{The line rms noise was measured from the stokes $I$ 
image cube using maser line free channels.}
\end{deluxetable}
     
\clearpage

\end{document}